\documentclass[preprint,review,12pt]{elsarticle}
\usepackage{graphicx}
\usepackage{epsfig}
\usepackage{natbib}
\usepackage{amssymb}
\usepackage{amsmath}
\usepackage{txfonts}

\def\be{\begin{equation}}
\def\ee{\end{equation}}

\journal{Astroparticle Physics}

\begin{document}

\begin{frontmatter}

\title{Space-time variation of the electron-to-proton mass ratio in a Weyl model}

\author[IFIBA]{Susana J. Landau \corref{cor1}\fnref{label1}} 
\ead{slandau@df.uba.ar} 
\address[IFIBA]{Instituto de F\'{\i}sica de Buenos Aires, Ciudad Universitaria - Pab. 1, 1428 Buenos Aires, Argentina} 
\author[FCAGLP]{Florencia A. Teppa Pannia }
\ead{flor@carina.fcaglp.unlp.edu.ar}
\address[FCAGLP]{Facultad de Ciencias Astron\'{o}micas y Geof\'{\i}sicas. Universidad Nacional de La Plata. Paseo del Bosque S/N 1900 La Plata, Argentina}
\author[UNAM]{Yuri Bonder}
\ead{yuri.bonder@nucleares.unam.mx}
\address[UNAM]{Instituto de Ciencias Nucleares, Universidad Nacional Aut\'{o}noma de M\'{e}xico, A. Postal 70-543, M\'{e}xico D.F. 04510, M\'{e}xico}
\author[UNAM,IAFE]{Daniel Sudarsky} 
\ead{sudarsky@nucleares.unam.mx}
\address[IAFE]{Instituto de Astronom{\'\i}a y F{\'\i}sica del Espacio, Casilla de Correos N 67, 1428 Buenos Aires, Argentina}

\cortext[cor1]{Corresponding author.} \fntext[label1]{Member of the Carrera del Investigador Cient\'{\i}fico y Tecnol\'ogico, CONICET.}




\begin{abstract}

Seeking a possible explanation for recent data indicating a space-time variation of the electron-to-proton mass ratio within the Milky Way, we consider a phenomenological model where the effective fermion masses depend on the local value of the Weyl tensor. We contrast the required values of the model's free parameters with bounds obtained from modern tests on the violation of the Weak Equivalence Principle and we find that these quantities are incompatible. This result indicates that the variation of nucleon and electron masses through a coupling with the Weyl tensor is not a viable model.

\end{abstract}

\end{frontmatter}

\section{Introduction}
\label{Intro}

The search for space-time dependence of fundamental constants plays a fundamental role in the continuous efforts to put in firmer empirical grounds our current physical theories and, at the same time, explore the possibilities of exotic physics that might become manifest trough small deviations. The experimental research can be grouped into astronomical and local methods. The latter ones include geophysical methods such as the natural nuclear reactor that operated about $1.8\times 10^9$ years ago in Oklo, Gabon \citep{DD96,Petrov06,Gould06}, the analysis of natural long-lived $\beta$ decays in geological minerals and meteorites \citep{Olive04b} and laboratory measurements such as comparisons of rates between clocks with different atomic numbers \citep{PTM95,Sortais01,Marion03,Bize03,Fischer04,Peik04}. The astronomical methods are based mainly on the analysis of high-redshift quasar absorption systems. Most of the reported data are, as expected, consistent with null variation of fundamental constants. Nevertheless, there are reports of intriguing results. For instance \citet{Webb99} and \citet{Murphy03b} have reported observations made with the Keck telescope which suggest a smaller value of the fine structure constant ($\alpha$) at high redshift as compared with its local value. However, an independent analysis performed with VLT/UVES data gave null results \citep{Srianand04}. Furthermore, a recent analysis using VLT/UVES data suggests also a variation in $\alpha$ but in the opposite sense, that is, $\alpha$ appears to be larger in the past \citep{Murphy10}. The discrepancy between Keck/HIRES and VLT/UVES is yet to be resolved. In particular, the two studies rely on data from different telescopes observing different hemispheres and it was pointed out that the Keck/Hires and VLT/UVES observations can be made consistent in the case where the fine structure constant is spatially varying \citep{Murphy10}. 

Focusing on a different quantity, observations of molecular hydrogen in quasar absorption systems can be used to set constraints on the electron-to-proton mass ratio $\mu \equiv m_e/m_p$ at high redshift \citep{King08, Thompson09,Malec10}, while the present value of $\mu$ can be constrained using comparisons of different transitions in atomic clocks \cite{Bize03,Fischer04,Peik04}. Furthermore, the observed temperature isotropy of microwave background radiation can be used to set bounds on the spatial variation of $\mu$ at extragalactic scales \citep{Barrow05}. Surprisingly, a recent analysis of ammonia spectra in the Milky Way suggests a spatial variation of $\mu$ \citep{MLK09,Levshakov10b,Levshakov10}. The study, comparing the spectral lines of the ammonia inversion transition and rotational transitions of other molecules with different sensitivities to the parameter $\mu$, finds a statistically significant velocity offset that when interpreted in terms of a variation in $\mu$ gives $\Delta \mu/\mu = (2.2 \pm 0.7) \times 10^{-8}$. This will be the focus of the present paper. If we assume that the latter is not the result of some fluke and systematic experimental error, and thus take the result quite seriously, we are naturally led to the following question: What would be the simplest modification of our present physical theories that might account for such phenomena? One of the simplest possibilities one can think of is that the effective value of the coupling constants changes with space-time location. In this sense we note that, within the context of theories that are at the fundamental level background independent, the study of possible space-time dependence of fundamental constants is often considered as equivalent to the search for the existence of dynamical fields which couple to the gauge fields and/or to ordinary matter in ways that mimic the ordinary coupling constants.

There have been several proposals along those lines with various different motivations. Some of them arise from proposals for basic theories that arise in the search for unification of the four fundamental laws of physics such as string-derived field theories \citep{Wu86,Maeda88,Barr88,DP94,DPV2002a,DPV2002b}, related brane-world theories \citep{Youm2001a,Youm2001b,branes03a,branes03b} and Kaluza-Klein theories \citep{Kaluza,Klein,Weinberg83,GT85,Weson97}. There are also phenomenological models where a scalar field $\phi$ couples to the Maxwell tensor $F_{\mu\nu}$ and are characterized by Lagrangian density terms such as $- B_F(\phi) F_{\mu\nu}F^{\mu\nu}/4$ \citep{Bekenstein82,BSM02,OP02,Wetterich03,Dent09} and/or to the matter fields $\Psi_i$ ($i$ labeling the field flavor) as $ B_i(\phi) \bar \Psi_i {\Psi}_i $ \citep{OP02,KW04a,Brax04,BM05,OP08,Wetterich03,Dent09}. If these terms were added to the standard Lagrangian densities, it is quite clear that, in the first case, it would result in something like an effective fine structure constant $1/ e^2_{effective} = 1/ e^2 + B_F(\phi)$ while in the second case the effective masses of the elementary particles would be given by $m_i^{eff} = m_i+ B_i(\phi)$. Then, if the field took space-time dependent values, a feature that often requires the new field to be quite light so its value is not too rigidly tied to the minima of any self-interaction potential, then the effective fine structure constant and/or effective masses might look space-time dependent. This part of the story is quite clear, however, one can not focus on just this aspect of the theory when considering it. In fact, it is often the case that the most important bounds on the theory do not arise from the direct search for this dependence but from the effects of the direct exchange of quanta of this putative field would have on the behavior of ordinary matter. The fact that the scalar field must be light, as we have just described, indicates that this quanta exchange would not be drastically suppressed by a large mass in its propagator \citep{OscillatingBounds}. This generically leads to modification of the free fall and very often to signals that would mimic violations of the weak equivalence principle (WEP). 
 
In fact, the connection between the theories involving spacetime dependence of coupling constants and the WEP was recognized already in the 1960s by Dicke \citep{dewitt1964relativity} (see page 163). More recently, \citet{Damour09} noted the extreme difficulty in mimicking the behavior of the mass of an object with any coupling that is not the gravitational one. This means that if the modified theory is required to be covariant, then the only possibility to ensure an exact compliance with the WEP is essentially to restrict the new fields to couple to matter in the same way as gravity, and this is impossible if we want the effective coupling constants to be scalar functions of some new fields. Consequently, one finds that the two issues are generically considered simultaneously in attempts to deal with possible variations of fundamental constants \citep{DP94,DPV2002a,DPV2002b,Bekenstein82,Barr88,Weson97,Scoccola07,BM05,OP02,Wetterich03,Dent09}. Thus, it is clear that we must face this connection in any attempt to deal with the problem at hand, and that in so doing we must consider the most modern and stringent bounds that are currently available on the possible violations of the WEP.

The basic idea of this manuscript is that rather than considering new fields which play the role of modifying the effective value of the fundamental constants, we can introduce non-standard aspects of well known fields in order to play that role. The long range fields in nature are the electromagnetic and gravitational ones. The use of the former in the desired context does not seem as a promising possibility because, for one it is very well understood and tested over very wide class of regimes, even at the quantum level (\textit{i.e.}, QED), and its enormous strength implies that any small modification would have very noticeable effects. The latter, on the other hand, is still far from being well understood (particularly its quantum aspects), and secondly, it seems conceivable that an exotic type of its coupling to matter might exist without having been detected so far. Considerations along these lines have led to proposals where the curvature of space-time might affect the propagation of matter fields in rather unusual ways \citep{QGP}, which might be viewed as violating of the strict letter of the equivalence principle without destroying overall general covariance\footnote{The proposals are motivated by searches for possible granular structure of space-time which might conflict with the ultra-locality that is implicitly assumed in the latter principle.}. In this manuscript we will explore this issue and show that, despite this early optimistic assessment, the value of the model's parameters needed to explain the observed variation of $\mu$ in the Milky Way can be ruled out by the bounds on those parameters emerging from experimental tests of the WEP. 

The paper is organized as follows. In section \ref{datos} we discuss the astronomical data that suggest a variation of $\mu$ in the Milky Way. In Section \ref{models} we describe a general theoretical model with an ``exotic'' coupling between matter and gravitation; in section \ref{ginv} we present a subclass of models where gauge invariance is preserved. Section \ref{fits} is devoted to determine the value of a combination of the free parameters for the models presented in section \ref{models} as implied by the astronomical data discussed in section \ref{datos}. In section \ref{fits2} we obtain bounds on another combination of the free parameters of the models using the latest tests of the WEP. We end with a brief discussion and some conclusive remarks in section \ref{discusion}.

\section{Data discussion}\label{datos}

Astronomical spectroscopy can probe physical constants which describe atomic and molecular discrete spectra. It is the case of the electron-to-proton mass ratio $\mu$. Recently, \citet{Levshakov10b,Levshakov10} reported new bounds on $\mu$ obtained through the ammonia method. Previous bounds with the same method were obtained by \citet{MLK09}. The method consists in comparing the observed frequency of the ${\rm NH_3}$ inversion transition with a suitable rotational frequency of another molecule arising co-spatially with ammonia. In particular, the authors used precise molecular lines observed in Milky Way cold dark clouds to compare the apparent radial velocity for the ${\rm NH_3}$ inversion transition, $V_{inv}$, with the apparent radial velocity, $V_{rot}$, for rotational transitions in ${\rm HC_3N}$ and ${\rm N_2H^+}$ arising from the same molecular cloud. The method provides a relation between the shift radial velocity and the relative variation of $\mu$:
\begin{equation} 
\frac{\Delta \mu}{\mu}=0.289\frac{V_{rot}-V_{inv}}{c}\equiv 0.289\frac{\Delta V}{c}.
\end{equation} 

In the first report \citep{Levshakov10}, three radio telescopes were used to obtain the data: 32-m Medicina, 100-m Effelsberg and 45-m Nobeyama. The authors found several problems in treating the data: i) not all the molecular profiles can be described adequately with a single component Gaussian model and ii) molecular cores are not ideal spheres and being observed at higher angular resolutions exhibit frequently complex substructures. The line profiles may be asymmetric due to non-thermal bulk motions. Therefore the authors selected 23 pair molecular lines from the initially 55 molecular pairs observed. The weighed mean and errors as well as the robust M-estimate of the mean reported for the 23 data, for 100-m Effelsberg and 45-m Nobeyama data sets are shown in table \ref{velocidad}. In order to check if the weighed mean is a representative value of the mean value of each data set of table \ref{velocidad} we have calculated $\chi^2 = \sum_i p_i (x_i-x_W)^2/\sum_j p_j$ ($x_W$ is the weighed mean, $p_i = 1/\sigma_i^2$ and $\sigma_i$ refers to the $1 \sigma$ error reported in \citet{Levshakov10}) and compared it with the expected value of $\chi^2$ for a Gaussian distribution. In all cases the obtained value of $\chi^2$ is large compared with the expected value of $\chi^2$ for a normal (Gaussian) distribution. Therefore, the distribution of the data does not seem to be normal (or Gaussian). The authors of \citet{Levshakov10} have also calculated the robust M-estimate of the mean. From table \ref{velocidad} it follows that there is a significant difference between the Nobeyama weighed mean and robust M-estimator, while for the complete data set and Effelsberg data set both estimators are consistent. On the other hand, although there is a good agreement between the robust M-estimate of the three data sets, the Nobeyama data set has a larger systematic error due to the lower accuracy at the rest frequencies of the $\rm{N_2^+ H}$ transition. Furthermore, from the $23$ sources of the complete data set, $2$ are in common for the $3$ telescopes, and $3$ are in common for Effelsberg and Nobeyama. \citet{Levshakov10} treated them as independent \textit{observations} because the data was obtained with different instruments, \textit{so such data points have different systematic errors}; however, these are not independent as measurements of the physical quantity.

In a recent paper, the authors map four molecular cores selected from their previous sample in order to estimate systematic effects in the velocity offset and to check the reproducibility of the velocity offset on the year-to-year time base \citep{Levshakov10b}. Observations were performed with the 100-m Effelsberg telescope. In two cores the velocity offset can be explained by the observed kinematic structure. In the other two cores, they obtain a statistically significant positive velocity offset which is shown in table \ref{velocidad}. The differences between densities in clouds are not significant, thus the Effelsberg robust M-estimate reported in \citet{Levshakov10} is statistically more significant than the value reported in \citet{Levshakov10b}. We will consider the value of the Effelsberg robust M-estimate reported in \citet{Levshakov10} as the final value of the velocity offset between rotational and inversion transitions. However, it should be noted that taking the other values reported in the paper or taking the value reported in \citet{Levshakov10b} will not produce a significant change in the conclusions of the theoretical models we are testing in this paper. 
\begin{table}[!ht]

\begin{center}
\renewcommand{\arraystretch}{1.5}
\caption{Results for the relative radial velocity for all the data, only data obtained with Effelsberg radio telescope and only data obtained with Nobeyama radio telescope \citep{Levshakov10}. Entry name Effelsberg (2) refers to results reported in \citet{Levshakov10b}. $\left< \Delta W \right>_W$ refers to the weighed mean and corresponding error, $\left< \Delta V\right>_M$ refers to the robust M-estimate of the mean. All velocities are in units of \rm{m/s}.}
\label{velocidad}
\begin{tabular}{lcc}
\hline
\hline Data & $\left< \Delta V \right>_W$&$\left< \Delta V \right>_M$ \\ \hline 
all&$20.7 \pm 3 $& $21.5 \pm 2.8$ \\ 
Effelsberg& $23.0 \pm 3.1$ &$23.2 \pm 3.8$\\ 
Nobeyama &$14.0 \pm 7.2$ & $22.9 \pm 4.2$ \\ 
Effelsberg (2)& & $26.9 \pm 4.2$ \\
\hline
\end{tabular}
\end{center}
\end{table}

\section{The Weyl Model}
\label{models}

The basic idea of the Weyl models involves considering the effect described in the previous section as due to a non-minimal and rather exotic coupling of the matter fields with gravity through the Weyl tensor. At first sight this may seem as an unnatural proposal as gravity is usually neglected in these regimes and, furthermore, one generally does not feel that there might be a fundamental reason to couple gravity with matter fields in exotic ways. However, in contrast to other models \citep{KW04a,Brax04,MS07,OP08}, in this scheme there is no need to invoke new unobserved dynamical fields, or non-dynamical fields that break Poincar\'e invariance, or other such problems, which in our opinion are ruled out by the analysis of its consequences on virtual particles \citep{Collins04, Collins06}. Moreover, as described in \citet{QGP}, Bonder and Sudarsky \cite{BS08,BS09,BS10}, the generic view we adopt in considering these sort of models is that gravity as ``the curvature of a manifold'' is only an effective description of more fundamental degrees of freedom (from an unknown quantum theory), and thus, the unnaturalness of the coupling terms is tied to the need to use a metric description rather than the still unknown language of the quantum gravity theory.

The task is to find a way in which gravity and matter could interact in a phenomenological level causing the described change in the matter's observed mass. We study fermionic matter fields $\Psi_i$ where $i$ labels the field's flavor. The usual mass term in this case is $m_i \bar{\Psi}_i \Psi_i$, therefore, in order to explain the observations we need to replace $m_i$ by some scalar depending on the gravitational environment. This should be implemented by 
\begin{equation}
m_i \rightarrow m_i \left[1+\xi_i f\left(\frac{R}{\Lambda^2}\right)\right],
\end{equation}
where $\xi_i$ are small phenomenological parameters which may be different for each flavor, $f$ is a function of the curvature tensor and $\Lambda$ is an energy scale. The first scalar function that comes to mind is the Ricci scalar $R$. However, the Ricci tensor at a space-time point $x$, and thus the Ricci scalar, is completely determined by the matter at the same point $x$, which implies that coupling $\Psi_i(x)$ with $R(x)$ is a self-coupling that, for phenomenological purposes, is not interesting. Thus, we should build $f$ with what is left when removing from the Riemann tensor the part determined by the Ricci tensor: The Weyl tensor, $W_{abcd}$. It is trivial to note that it is not possible to construct a Lorentz scalar out of one power of $W_{abcd}$, thus, the simplest scalar one can write is $f=\frac{W_{abcd} W^{abcd}}{\Lambda^4}\equiv\frac{W^2}{\Lambda^4}$. Note that $\Lambda^4$ has (mass) dimensions $4$ ($\hbar$ and $c$ are taken to be $1$) and thus $\xi_i$ are dimensionless parameters.

 

\subsection{Gauge invariance and the Weyl proposal}
\label{ginv}

We might be concerned that if the coupling to different fermion flavors are completely arbitrary, the new proposal conflicts with gauge invariance destroying the renormalizability of the theory. However, the fact that this modification is assumed to be tied to gravitational sector itself, known to be non-renormalizable\footnote{Nowadays this is simply taken as indication that a theory must be 
considered as an effective theory.}, places these concerns in the proper perspective. In addition, a simple restriction can remove this issue altogether: The quantity that we need to vary with location is the ratio of electron and proton masses $\mu$, and the two quantities have a rather different origin; one arising mainly a from the strong interaction of the quark and gluon constituents, while the other is due to the Yukawa term and the vacuum expectation value of the Higgs field. We can construct a rather simple theory with the required features by focusing on this difference. Namely, we assume that the exotic coupling to Weyl tensor enters into the theory only in the fermionic Yukawa terms as
\begin{equation}
{\cal L}_{Yukawa}=\sum_{a,b} \Gamma_{ab} (W^2) \bar{\psi}_{a}^{R} \Phi \Psi_b^{L} + h.c.
\end{equation}
where the sum is over the standard SU(2) left handed fermion field doublets $\Psi_b^{L}$ and right handed fermion filed singlets $\psi^R_b$ of the electroweak theory, $\Phi$ is the standard doublet Higgs doublet field, and $\Gamma_{ab}(W^2) $ are the Yukawa coefficients which are now considered to depend on the local value of the magnitude of Weyl tensor. Considering the lowest order of a Taylor expansion we can write $\Gamma_{ab} (W^2) = (1+{\xi}_{(ab)} W^2 /\Lambda^4 )\Gamma_{ab} (0)$, where the parameters $\xi_{ab}$ characterize the leading order corrections and, in general, depend on the flavor indices $a$ and $b$. As we only focus on the first generation and the quark sector is, in any event, quite different phenomenologically from the lepton sector due to the connection with the strong (or color) force, we ignore that possible dependence and simply write $\xi$. It is thus clear that as the Higgs field acquires a vacuum expectation value, the resulting fermionic mass matrices are multiplied by the factor $1+ \xi W^2/\Lambda^4$ and thus the lepton and quark masses appear in the effective Lagrangian as multiplied by that factor.
 
It is then evident that, as long as gravitation is considered as a fixed background and one does not attempt to introduce radiative corrections involving gravitons, the theory preserves the same renormalizability properties as the standard model. Of course, when attempting to include radiative corrections involving gravitons, the theory will encounter the usual problems facing the quantization of gravitation. On the other hand, the mass of any lepton appears in the low energy theory as 
\begin{equation}
m_l\rightarrow m_l \left(1+ \frac{\xi}{\Lambda^4}W^2 \right),
\end{equation}
while in the case of hadrons only the valence quark mass, which represents a rather small part of the total hadron mass, suffers such modification, thus we can expect that the change in the total hadron mass is much smaller, \textit{i.e.},
\begin{equation}
m_h\rightarrow m_h \left(1+ a_h \frac{\xi}{\Lambda^4}W^2 \right),
\label{hadron}
\end{equation}
where the parameter $ a_h$ is of order $10^{-3}$ (the valence quark proportion of the hadronic masses). In other words, we can take the electron's $\xi_e$ parameter to be identical to the fundamental parameter $\xi$, while the parameter corresponding to the proton $\xi_p$ is given by $ a_p \xi$ where $ a_p$ is of order $10^{-3}$. The model described in this subsection is less general than the case where the proton and electron coupling to the Weyl tensor are not related. In the following sections we consider the theoretical predictions of both models and compare it with astronomical and experimental data. We call model I to the model where the couplings of leptons and quarks are not related and model II to the case where the exotic coupling enters only in the Yukawa terms.
 
\section{Estimates on the Weyl model parameters from electron-to-proton mass ratio}\label{fits}

In a model where the interaction between gravity and matter at a phenomenological level is such as described in section \ref{models}, the low energy limit of the interaction Lagrangian density can be written as
\begin{equation}\label{lag}
{\cal L}_{int}= \frac{\xi_e}{M_P^4} m_e W^2 \bar \psi_e \psi_e
+\frac{\xi_p}{M_P^4} m_p W^2 \bar \psi_p \psi_p +\frac{\xi_n}{M_P^4} m_n W^2 \bar \psi_n \psi_n,
\end{equation}
where subscripts $e,p,n$ respectively refer to electrons, protons and neutrons and we set $\Lambda$ equal to the Planck scale $M_P = 1.22 \times 10^{19}$ GeV. Thus, the effective masses of the particles can be expressed as $m^{eff}_i = m_i\left(1 + \frac{\xi_i}{M_P^4} W^2\right)$. Note that we have restored in the notation $\xi_i$ the possible dependence of the parameter on the particle type. The point being, as discussed in the previous section, that even if $\xi$ had a single value throughout the first generation of quarks and leptons, the composite nature of the hadrons, as well as the gluon contribution to the mass of these particles, would make the effective value quite different from that corresponding to electrons. The difference between neutrons and protons can similarly be expected to arise as a result of the electromagnetic differences between them. 

According to the model we are proposing, the observable quantity is
\begin{equation}
\mu^{eff} \equiv \frac{m^{eff}_e}{m^{eff}_p}=\frac{m_e}{m_p}\left(\frac{1+\frac{\xi_e}{M_P^4}
W^2}{1+\frac{\xi_p}{M_P^4} W^2}\right) \simeq \frac{m_e}{m_p}\left(1+\alpha \frac{W^2}{M_P^4}\right),
\end{equation}
where in the last step we use the fact that $\xi_i/M_P^4$ are small and we define $\alpha\equiv \xi_e -\xi_p $ for model I and ${\alpha\equiv (1-a_h)\xi}$ for model II. Thus, the observed electron-to-proton mass ratio value in cold molecular clouds respect to the same value at Earth is
\begin{equation}\label{Delta mu eff}
\frac{\mu^{eff}_{cl}-\mu^{eff}_{\oplus}}{\mu^{eff}_{\oplus}} \simeq \frac{\alpha}{M_P^4} (W^2_{cl} -W^2_{\oplus}),
\end{equation}
where the subscripts $cl$ and $\oplus$ stand respectively for the interstellar clouds and the Earth.

In order to compute the Weyl tensor both, in the cloud and at the laboratory, we consider a sphere of constant density $\rho$ and radius $R$, surrounded by vacuum. Outside the sphere, namely at a distance $r\geq R$ from its center, we get $W^2=\frac{48 G^2 M^2}{r^6}$ where $G$ is Newton's constant and $M \equiv 4 \pi \rho R^3/3$ is the mass of the sphere. Due to the dependence of $W^2$ on $r$ we realize that we have to be careful while considering the contributions to $W^2$. For example, contributions from massive bodies near to the laboratory such as a wall may be greater than the contribution of the entire Earth. Therefore, we also consider  the contribution of the Earth and from walls of $4$ m height, $4$ m width and $0.5$ m depth made of cement and iron located $0.1$ m from the experiment.  The results of the calculations show that the contribution to $W^2$  of an iron wall   located very close to the experiment ($W^2_{I}= 1.8 \times 10^{-44} {\rm m^{-4}}$) is   greater than the effect of the Earth ($W^2_{\oplus} = 1.4 \times 10^{-44} {\rm m^{-4}}$)  or a cement wall ($W^2_{C} = 1.4 \times 10^{-44} {\rm m^{-4}}$) .
Given that we have not the exact details of the laboratory where the rest wavelengths are measured and since we are dealing with a positive detection of $\Delta\mu^{eff}$, we consider the iron wall to estimate a lower bound on $\alpha$. On the other hand, the contribution of the entire Earth can not be neglected and therefore we consider this contribution to obtain an upper bound on $\alpha$. Even though the wall is not a sphere, in the regime we are working on it is possible to use the approximation of linearized gravity where the superposition principle is valid. Thus, the contribution of the wall can be regarded as the sum of the contribution of a great number of spheres. We estimate that differences with the exact calculation may be at most of order $10$.


We model the interstellar clouds as spheres with constant density $\rho$ and radios $R$. In order to calculate $W^2$ we use the inner Schwarzschild solution, which leads to $W^{2}_{cl}=\frac{4^{6}\pi^{4}r^{4}\rho^4 G^{4}(3-8\pi
R^{2}\rho G)^{2}}{3(3-8\pi r^{2}\rho G)^2(3-2\pi \rho G(3r^{2}+R^{2}))^{2}}$. Taking for the clouds a mean density $\rho_{cl}= 1.5\times 10^{27}$ m$^{-4}$, a mean radius $R_{cl} = 0.052$ Pc \citep{MLK09} and at $r=R/2$ we obtain $W^{2}_{cl}\simeq 1.5 \times 10^{-178}\ \rm{m}^{-4}$.(We also calculated the value of $W^{2}_{cl}$ for $r=0$ and $r=R$ and the results do not differ significantly from the case $r=R/2$). Taking the value of Effelsberg robust mean discussed in section \ref{datos} we get
\begin{equation}\label{ineq alpha1}
1.22 \times 10^{36}\rm{m}^4 \le \frac{|\alpha|}{M_P^4} \le 1.57 \times 10^{36}\rm{m}^4
\end{equation}
On the other hand, if we consider the model described in section \ref{ginv}, where hadron masses suffer much smaller modifications than lepton masses, in order for the model to explain the observations described in section \ref{datos} the following condition must be fulfilled:
\begin{equation}\label{ineq alpha2}
1.22 \times 10^{36}\rm{m}^4 \le \frac{|\xi|}{M_P^4} \le 1.57 \times 10^{36}\rm{m}^4
\end{equation}
The next section is dedicated to study if the relations (\ref{ineq alpha1}) and (\ref{ineq alpha2}) are compatible tests of the WEP.

\section{Bounds from E\"otvos type experiments}\label{fits2}

The gravitational potential of an object composed of $N$ atoms with atomic number $Z$ and baryon number $B$ can be written by taking into account that the effective masses are modified in this model according to equation (\ref{lag}):
\begin{eqnarray}
V &=& N Z \frac{\xi_e}{M_P^4} m_e W^2 + N Z \frac{\xi_p}{M_P^4} m_p W^2 + N
(B-Z) \frac{\xi_n}{M_P^4} m_n W^2 \nonumber \\
&=& N \frac{\alpha'} {M_P^4} W^2 m_p,
\end{eqnarray}
where
\begin{equation}
\alpha' =Z \left(\xi_e \frac{m_e}{m_p} + \xi_p - \xi_n \frac{m_n}{m_p}
\right)+B\xi_n \frac{m_n}{m_p}.
\end{equation}
The force acting on a freely falling body of mass $M_b$ can be obtained from $\vec F= -\vec \nabla V$, thus, the respective acceleration is $\vec{a} = -\frac{N \alpha' m_p \vec{\nabla} W^2}{M_P^4 M_b}$. The differential acceleration of two bodies with different composition but the same number of atoms $N$ is $\Delta \vec{a}= -\frac{\vec{\nabla} W^2\ m_p (\delta_1\ \alpha_1' + \delta_2 \alpha_2')}{M_P^4}$, where we assume that the mass of the body can be expressed as $N m$ with $m$ the atomic mass of each body and we define
\begin{equation}
\alpha_1'\equiv \xi_e \frac{m_e}{m_p} + \xi_p - \xi_n \frac{m_n}{m_p}, \qquad 
\alpha_2'\equiv \xi_n \frac{m_n}{m_p},
\end{equation}
and also $\delta_1\equiv (Z_1/m_1)-(Z_2/m_2)$ and $\delta_2\equiv (B_1/m_1)-(B_2/m_2)$, the subscripts indicating which object we are considering. On the other hand, if hadron masses suffer smaller modifications as suggested in section \ref{ginv}, the prediction of the violation of WEP can be written in terms of a single parameter $\xi$: $\alpha_1'\equiv 10^{-3} \xi$, $\alpha_2'\equiv 10^{-3} \xi$.

Limits on violations of the WEP come from E\"otvos-Roll-Krotkov-Dicke and Braginsky-Pannov measurements of the differential acceleration of test bodies. The most stringent limits are obtained from measurements of differential acceleration towards the Sun. However, since the force resulting from Weyl models is a short range force, the relevant bounds to test such models are provided by measurements towards the Earth. In this kind of experiments a continuously rotating torsion balance instrument is used to measure the acceleration difference of test bodies with different composition. In table \ref{tabla1}, we summarize the current bounds considered in this paper as well as the composition of the test bodies in each case. All bounds were obtained from an experiment made at the University of Washington Nuclear Physics Laboratory. The authors of the experiment mention two sources for the relevant signals and in our case these would be the sources for $\vec{\nabla} W^2$, a hillside of $26$ m located closed to the laboratory (we estimate $1$ m) and a layer of cement blocks added to the wall of the laboratory \citep{Adelberger90}. Here we model these sources simply as spherical masses at a given distance. Barring some fortuitous cancellation among the various contributions in the laboratory due to their detailed location and geometry, these simplification cannot amount to more than a change by a geometrical factor of order one, which we ignore. Moreover, the expression for $\left|\vec{\nabla} W^2\right|$ reads $\left|\vec{\nabla} W^2\right|=\frac{288}{r^7} \left(\frac{4}{3}\pi G \rho R^3\right)^2$, and therefore the contribution from the hillside and cement layer can be estimated to be, respectively, $\left|\vec{\nabla} W^2_{\rm{H}}\right|= 7.9 \times 10^{-45} \rm{m}^{-5}$, and $\left|\vec{\nabla} W^2_{\rm{L}}\right|=2.9 \times 10^{-45} \rm{m}^{-5}$. Since both contribution are of the same order, we continue our analysis considering only the contribution of the hillside. In order to estimate the value of $\alpha'_1$ and $\alpha'_2$, we perform a least square minimization using the calculated expression for the acceleration and the data from table \ref{tabla1}. We obtain for model I:
\begin{eqnarray}
\frac{\alpha_1'}{M_P^4}&=&(3.5 \pm 7.8) \times 10^{13}\ \rm{m}^4,\\
\frac{\alpha_2'}{M_P^4}&=&(-0.7 \pm 2) \times 10^{15}\ \rm{m}^4.
\end{eqnarray}
Moreover, using the data from table \ref{tabla1} we get for the model II:
\begin{equation}
\frac{\xi}{M_P^4} = (1.7 \pm 6) \times 10^{16} \rm{m}^4.
\end{equation}
The contribution of other massive sources such as other hills close to the laboratory would give lower values of $\left|\vec{\nabla} W^2\right|$ resulting in higher values for $\xi$. However, we cannot ignore the hill of $26$ m and the layer of cement mentioned by \citet{Adelberger90} and therefore the limit on $\xi$ is not a lower bound in the sense that it has been discussed in section \ref{fits} for the astronomical data.

\begin{table*}[!ht]
\renewcommand{\arraystretch}{1.3}
\caption{Bounds on the equivalence principle considered in this paper.$(B/{\rm m})_i$ and $M({\rm g})$ refer to the baryon number and total mass of the test bodies }
\label{tabla1}
\begin{center}
\begin{tabular}{ccccccc}
\hline
\hline
$\Delta a \,\, (\rm{m}^{-1})$&$Z_1$&$Z_2$&$(B/{\rm m})_1$&$(B/{\rm m})_2$&$M
({\rm g})$ & Reference\\\hline
 $(2.5 \pm 7.23) \times 10^{-31}$&$ 4$&$13$&$0.998648$&$1.000684$&$10$
&\cite{Su94}\\
 $(5.3 \pm 6.45) \times 10^{-31}$&$4$&$29$&$0.998648$&$1.001117$&$10$
&\cite{Su94}\\
 $(0.67 \pm 3.45) \times 10^{-32}$&$4$&$22$&$0.99868$&$1.001077$&$4.8$ &
\cite{Schlamminger08} \\
 $(2.3 \pm 2.3) \times 10^{-30}$&$4$&$13$&$0.998648$&$1.000684$&$10$
&\cite{Adelberger90}\\
 $(0.9 \pm 1.89) \times 10^{-30}$&$4$&$29$&$0.998648$&$1.001117$&$10 $ &
\cite{Adelberger90}\\
 $(1.11 \pm 3.11) \times 10^{-32}$&$29$&$82$&$1.0011166$&$1.0001694$&$10$&
\cite{Smith00}\\\hline
\end{tabular}
\end{center}
\end{table*}

\section{Discussion and Conclusions}
\label{discusion}

We have considered a model where a non-minimal coupling of Weyl tensor to matter would result in an effective mass for fermionic fields which would be space-time dependent. The model might be considered as a possible explanation for the recent reported observations of a space-time variation of the electron-to-proton mass ratio in \cite{MLK09,Levshakov10b,Levshakov10}. We have developed the model in some detail and extracted the range of values for a combination of the parameters which would be necessary to account for the ``exotic'' observation. We have also considered the constrains on the model that arise from consideration of precision tests of the WEP and used some of the most modern relevant data to contrast the two results.

In section \ref{fits} we analyzed the variation of the effective electron-to-proton mass ratio between the environments corresponding to the Earth's surface and a molecular cloud and showed that consistency with the reported data requires $\alpha = \xi_e - \xi_p \sim 10^{36}\  \rm{m}^4\  M_P^4$. On the other hand, the result of the statistical analysis performed in section \ref{fits2} with bounds on the WEP constrain the value of $\alpha_1'=\xi_e m_e/m_p + \xi_p - \xi_n m_n/m_p$ to be of order $10^{13}\  \rm{m}^4\  M_P^4$. Here one might be inclined to note that the two situations are sensitive to slightly different combination of the fundamental parameters, and is thus conceivable that $\alpha$ might be as large as required to account for the astronomical observations while $\alpha_1'$ is as small as needed to conform with the laboratory bounds. We view such possibility as very unlikely as it would imply that the particular choice of materials compared in the the laboratory tests were coincidentally those for which the signal resulting from the generic couplings happened to cancel out almost exactly (at the level of one part in $10^{17}$). Moreover, as similar tests with slightly lower precision do exist for other materials, taking this line of reasoning would only lead to a reduction by at most a couple of orders of magnitude in the constraint, something which would still be sufficient to rule the model out.

If we consider the model described in subsection \ref{ginv}, consistency with the astronomical data requires $|\xi| \sim 10^{36}\  \rm{m}^4\  M_P^4$ whereas the result of the statistical analysis performed in section \ref{fits2} with bounds on the WEP constrain the value of $\xi $ to be of order $10^{16}\  \rm{m}^4\  M_P^4$. We thus have found that barring some miraculous cancellation or some unnatural fine tuning of the experimental conditions and/or of the model, the two sets are incompatible, and that a model where the variation of the electron, proton and neutron effective mass is driven by the scalar magnitude of the Weyl tensor can not account for the experimental constraints and the observational data and should be also ruled out.

If the observations of \cite{MLK09,Levshakov10b,Levshakov10} were to be further confirmed and the evidence for a change in the value of the electron-to-proton mass ratio became incontrovertible, one would need some different sort of explanation, however, due to the connection between space-time dependency of parameters and the couplings of ordinary matter with dynamical fields, which appears inherent of background independent theories, it seems very unlikely that one might find a model where the bounds imposed by tests of the WEP would not be of great relevance and impact. Nonetheless, we should stress our belief that this kind of models should be further explored, not only as potential explanatory grounds for atypical observations, but also as leading to robust constraints on the possible nontrivial couplings of matter and gravitation. 

\section*{{\bf Acknowledgments}}
We gladly acknowledge very useful discussions with E. Fischbach and P. Molaro. Y.B. and D.S. acknowledge support from PAPPIT project IN 119808 and CONACYT project No 101712. D.S. was supported in part by sabbatical fellowships from CONACYT and DGAPA-UNAM and the hospitality of the IAFE. S. J. L. and F.A. T. P. are supported by PICT 2007-02184 from Agencia Nacional de Promoci\'{o}n Cient\'{i}fica y Tecnol\'{o}gica, Argentina and by PIP N 11220090100152 from Consejo Nacional de Investigaciones Cient\'{\i}ficas y T\'{e}cnicas, Argentina. 

\bibliography{bibliografia6}

\begin{thebibliography}{60}
\expandafter\ifx\csname natexlab\endcsname\relax\def\natexlab#1{#1}\fi
\providecommand{\bibinfo}[2]{#2}
\ifx\xfnm\relax \def\xfnm[#1]{\unskip,\space#1}\fi
\bibitem[{{Adelberger et al}(1990)}]{Adelberger90}
\bibinfo{author}{E.G. {Adelberger et al}}, \bibinfo{journal}{Phys. Rev. D}
  \bibinfo{volume}{42} (\bibinfo{year}{1990}) \bibinfo{pages}{3267}.
\bibitem[{{Sortais}~et al(2001)}]{Sortais01}
\bibinfo{author}{Y.~{Sortais}~et al}, \bibinfo{journal}{Physica Scripta Volume
  T} \bibinfo{volume}{95} (\bibinfo{year}{2001}) \bibinfo{pages}{50}.
\bibitem[{{Barr} and {Mohapatra}(1988)}]{Barr88}
\bibinfo{author}{S.M. {Barr}}, \bibinfo{author}{P.K. {Mohapatra}},
  \bibinfo{journal}{Phys. Rev. D} \bibinfo{volume}{38} (\bibinfo{year}{1988})
  \bibinfo{pages}{3011}.
\bibitem[{{Barrow}(2005)}]{Barrow05}
\bibinfo{author}{J.D. {Barrow}}, \bibinfo{journal}{Phys. Rev. D}
  \bibinfo{volume}{71} (\bibinfo{year}{2005}) \bibinfo{pages}{083520}.
\bibitem[{{Barrow} and {Magueijo}(2005)}]{BM05}
\bibinfo{author}{J.D. {Barrow}}, \bibinfo{author}{J.~{Magueijo}},
  \bibinfo{journal}{Phys. Rev. D} \bibinfo{volume}{72} (\bibinfo{year}{2005})
  \bibinfo{pages}{043521}.
\bibitem[{{Barrow et al}(2002)}]{BSM02}
\bibinfo{author}{J.D. {Barrow et al}}, \bibinfo{journal}{Phys. Rev. D}
  \bibinfo{volume}{65} (\bibinfo{year}{2002}) \bibinfo{pages}{063504}.
\bibitem[{Bekenstein(1982)}]{Bekenstein82}
\bibinfo{author}{J.D. Bekenstein}, \bibinfo{journal}{Phys. Rev. D}
  \bibinfo{volume}{25} (\bibinfo{year}{1982}) \bibinfo{pages}{1527}.
\bibitem[{{Bize et al}(2003)}]{Bize03}
\bibinfo{author}{S.~{Bize et al}}, \bibinfo{journal}{Phys. Rev. Lett.}
  \bibinfo{volume}{90} (\bibinfo{year}{2003}) \bibinfo{pages}{150802}.
\bibitem[{{Bonder} and {Sudarsky}(2008)}]{BS08}
\bibinfo{author}{Y.~{Bonder}}, \bibinfo{author}{D.~{Sudarsky}},
  \bibinfo{journal}{Class. Quantum Grav.} \bibinfo{volume}{25}
  (\bibinfo{year}{2008}) \bibinfo{pages}{105017}.
\bibitem[{Bonder and Sudarsky(2009)}]{BS09}
\bibinfo{author}{Y.~Bonder}, \bibinfo{author}{D.~Sudarsky},
  \bibinfo{journal}{Rep. Math. Phys.} \bibinfo{volume}{64}
  (\bibinfo{year}{2009}) \bibinfo{pages}{169}.
\bibitem[{{Bonder} and {Sudarsky}(2010)}]{BS10}
\bibinfo{author}{Y.~{Bonder}}, \bibinfo{author}{D.~{Sudarsky}},
  \bibinfo{journal}{AIP Conference Series} \bibinfo{volume}{1256}
  (\bibinfo{year}{2010}) \bibinfo{pages}{157}.
\bibitem[{{Brax et al}(2003)}]{branes03b}
\bibinfo{author}{P.~{Brax et al}}, \bibinfo{journal}{Astrophysics and Space
  Science} \bibinfo{volume}{283} (\bibinfo{year}{2003}) \bibinfo{pages}{627}.
\bibitem[{{Brax et al}(2004)}]{Brax04}
\bibinfo{author}{P.~{Brax et al}}, \bibinfo{journal}{Phys. Rev. D}
  \bibinfo{volume}{70} (\bibinfo{year}{2004}) \bibinfo{pages}{123518}.
\bibitem[{{Collins et al}(2004)}]{Collins04}
\bibinfo{author}{J.~{Collins et al}}, \bibinfo{journal}{Phys. Rev. Lett.}
  \bibinfo{volume}{93} (\bibinfo{year}{2004}) \bibinfo{pages}{191301}.
\bibitem[{{Collins et al}(2009)}]{Collins06}
\bibinfo{author}{J.~{Collins et al}}, in: \bibinfo{editor}{D.~Oriti} (Ed.),
  \bibinfo{booktitle}{Approaches to Quantum Gravity: Toward a New Understanding
  of Space, Time and Matter}, \bibinfo{publisher}{Cambridge University Press},
  \bibinfo{year}{2009}.
\bibitem[{{Corichi} and {Sudarsky}(2005)}]{QGP}
\bibinfo{author}{A.~{Corichi}}, \bibinfo{author}{D.~{Sudarsky}},
  \bibinfo{journal}{Int. J. of Mod. Phys. D} \bibinfo{volume}{14}
  (\bibinfo{year}{2005}) \bibinfo{pages}{1685}.
\bibitem[{{Damour}(2009)}]{Damour09}
\bibinfo{author}{T.~{Damour}}, \bibinfo{journal}{Space Science Reviews}
  \bibinfo{volume}{148} (\bibinfo{year}{2009}) \bibinfo{pages}{191}.
\bibitem[{{Damour} and {Dyson}(1996)}]{DD96}
\bibinfo{author}{T.~{Damour}}, \bibinfo{author}{F.~{Dyson}},
  \bibinfo{journal}{Nucl.Phys.B} \bibinfo{volume}{480} (\bibinfo{year}{1996})
  \bibinfo{pages}{37}.
\bibitem[{{Damour} and {Polyakov}(1994)}]{DP94}
\bibinfo{author}{T.~{Damour}}, \bibinfo{author}{A.M. {Polyakov}},
  \bibinfo{journal}{Nucl.Phys.B} \bibinfo{volume}{423} (\bibinfo{year}{1994})
  \bibinfo{pages}{532}.
\bibitem[{{Damour et al}(2002{\natexlab{a}})}]{DPV2002a}
\bibinfo{author}{T.~{Damour et al}}, \bibinfo{journal}{Phys. Rev. Lett.}
  \bibinfo{volume}{89} (\bibinfo{year}{2002}{\natexlab{a}})
  \bibinfo{pages}{081601}.
\bibitem[{{Damour et al}(2002{\natexlab{b}})}]{DPV2002b}
\bibinfo{author}{T.~{Damour et al}}, \bibinfo{journal}{Phys. Rev. D}
  \bibinfo{volume}{66} (\bibinfo{year}{2002}{\natexlab{b}})
  \bibinfo{pages}{046007}.
\bibitem[{{Dent et al}(2009)}]{Dent09}
\bibinfo{author}{T.~{Dent et al}}, \bibinfo{journal}{JCAP} \bibinfo{volume}{1}
  (\bibinfo{year}{2009}) \bibinfo{pages}{38}.
\bibitem[{DeWitt and DeWitt(1964)}]{dewitt1964relativity}
\bibinfo{author}{C.~DeWitt}, \bibinfo{author}{B.S. DeWitt},
  \bibinfo{title}{Relativity, Groups and Topology (Houches Lecture)},
  \bibinfo{publisher}{Gordon \& Breach Science Publishers},
  \bibinfo{year}{1964}.
\bibitem[{{Fischer et al}(2004)}]{Fischer04}
\bibinfo{author}{M.~{Fischer et al}}, \bibinfo{journal}{Phys. Rev. Lett.}
  \bibinfo{volume}{92} (\bibinfo{year}{2004}) \bibinfo{pages}{230802}.
\bibitem[{{Gleiser} and {Taylor}(1985)}]{GT85}
\bibinfo{author}{M.~{Gleiser}}, \bibinfo{author}{J.G. {Taylor}},
  \bibinfo{journal}{Phys. Rev. D} \bibinfo{volume}{31} (\bibinfo{year}{1985})
  \bibinfo{pages}{1904}.
\bibitem[{{Gould et al}(2006)}]{Gould06}
\bibinfo{author}{C.R. {Gould et al}}, \bibinfo{journal}{Phys. Rev. C}
  \bibinfo{volume}{74} (\bibinfo{year}{2006}) \bibinfo{pages}{024607}.
\bibitem[{Kaluza(1921)}]{Kaluza}
\bibinfo{author}{T.~Kaluza}, \bibinfo{journal}{Sitzungber. Preuss. Akad.
  Wiss.K} \bibinfo{volume}{1} (\bibinfo{year}{1921}) \bibinfo{pages}{966}.
\bibitem[{{Khoury} and {Weltman}(2004)}]{KW04a}
\bibinfo{author}{J.~{Khoury}}, \bibinfo{author}{A.~{Weltman}},
  \bibinfo{journal}{Phys. Rev. Lett.} \bibinfo{volume}{93}
  (\bibinfo{year}{2004}) \bibinfo{pages}{171104}.
\bibitem[{{King et al}(2008)}]{King08}
\bibinfo{author}{J.A. {King et al}}, \bibinfo{journal}{Phys. Rev. Lett.}
  \bibinfo{volume}{101} (\bibinfo{year}{2008}) \bibinfo{pages}{251304}.
\bibitem[{Klein(1926)}]{Klein}
\bibinfo{author}{O.~Klein}, \bibinfo{journal}{Z. Phys.} \bibinfo{volume}{37}
  (\bibinfo{year}{1926}) \bibinfo{pages}{895}.
\bibitem[{{Levshakov et al}(2010{\natexlab{a}})}]{Levshakov10b}
\bibinfo{author}{S.A. {Levshakov et al}}, \bibinfo{journal}{A\&A}
  \bibinfo{volume}{524} (\bibinfo{year}{2010}{\natexlab{a}})
  \bibinfo{pages}{A32}.
\bibitem[{{Levshakov et al}(2010{\natexlab{b}})}]{Levshakov10}
\bibinfo{author}{S.A. {Levshakov et al}}, \bibinfo{journal}{A\&A}
  \bibinfo{volume}{512} (\bibinfo{year}{2010}{\natexlab{b}})
  \bibinfo{pages}{A44}.
\bibitem[{{Maeda}(1988)}]{Maeda88}
\bibinfo{author}{K.~{Maeda}}, \bibinfo{journal}{Mod. Phys. Lett. A}
  \bibinfo{volume}{3} (\bibinfo{year}{1988}) \bibinfo{pages}{243}.
\bibitem[{{Malec et al }(2010)}]{Malec10}
\bibinfo{author}{A.L. {Malec et al }}, \bibinfo{journal}{Mon. Not. R. Astron.
  Soc.} \bibinfo{volume}{403} (\bibinfo{year}{2010}) \bibinfo{pages}{1541}.
\bibitem[{{Marion et al}(2003)}]{Marion03}
\bibinfo{author}{H.~{Marion et al}}, \bibinfo{journal}{Phys. Rev. Lett.}
  \bibinfo{volume}{90} (\bibinfo{year}{2003}) \bibinfo{pages}{150801}.
\bibitem[{{Molaro et al}(2009)}]{MLK09}
\bibinfo{author}{P.~{Molaro et al}}, \bibinfo{journal}{Nucl.Phys.B Proc.
  Suppl.} \bibinfo{volume}{194} (\bibinfo{year}{2009}) \bibinfo{pages}{287}.
\bibitem[{{Mota} and {Shaw}(2007)}]{MS07}
\bibinfo{author}{D.F. {Mota}}, \bibinfo{author}{D.J. {Shaw}},
  \bibinfo{journal}{Phys. Rev. D} \bibinfo{volume}{75} (\bibinfo{year}{2007})
  \bibinfo{pages}{063501}.
\bibitem[{{Murphy et al}(2003)}]{Murphy03b}
\bibinfo{author}{M.T. {Murphy et al}}, \bibinfo{journal}{Mon. Not. R. Astron.
  Soc.} \bibinfo{volume}{345} (\bibinfo{year}{2003}) \bibinfo{pages}{609}.
\bibitem[{{Olive} and {Pospelov}(2002)}]{OP02}
\bibinfo{author}{K.A. {Olive}}, \bibinfo{author}{M.~{Pospelov}},
  \bibinfo{journal}{Phys. Rev. D} \bibinfo{volume}{65} (\bibinfo{year}{2002})
  \bibinfo{pages}{085044}.
\bibitem[{{Olive} and {Pospelov}(2008)}]{OP08}
\bibinfo{author}{K.A. {Olive}}, \bibinfo{author}{M.~{Pospelov}},
  \bibinfo{journal}{Phys. Rev. D} \bibinfo{volume}{77} (\bibinfo{year}{2008})
  \bibinfo{pages}{043524}.
\bibitem[{{Olive et al}(2004)}]{Olive04b}
\bibinfo{author}{K.A. {Olive et al}}, \bibinfo{journal}{Phys. Rev. D}
  \bibinfo{volume}{69} (\bibinfo{year}{2004}) \bibinfo{pages}{027701}.
\bibitem[{{Overduin} and {Wesson}(1997)}]{Weson97}
\bibinfo{author}{J.M. {Overduin}}, \bibinfo{author}{P.S. {Wesson}},
  \bibinfo{journal}{Phys. Rep.} \bibinfo{volume}{283} (\bibinfo{year}{1997})
  \bibinfo{pages}{303}.
\bibitem[{{Palma et al}(2003)}]{branes03a}
\bibinfo{author}{G.A. {Palma et al}}, \bibinfo{journal}{Phys. Rev. D}
  \bibinfo{volume}{68} (\bibinfo{year}{2003}) \bibinfo{pages}{123519}.
\bibitem[{{Peik et al}(2004)}]{Peik04}
\bibinfo{author}{E.~{Peik et al}}, \bibinfo{journal}{Phys. Rev. Lett.}
  \bibinfo{volume}{93} (\bibinfo{year}{2004}) \bibinfo{pages}{170801}.
\bibitem[{{Petrov et al}(2006)}]{Petrov06}
\bibinfo{author}{Y.V. {Petrov et al}}, \bibinfo{journal}{Phys. Rev. C}
  \bibinfo{volume}{74} (\bibinfo{year}{2006}) \bibinfo{pages}{064610}.
\bibitem[{{Prestage et al}(1995)}]{PTM95}
\bibinfo{author}{J.D. {Prestage et al}}, \bibinfo{journal}{Phys. Rev. Lett.}
  \bibinfo{volume}{74} (\bibinfo{year}{1995}) \bibinfo{pages}{3511}.
\bibitem[{{Schlamminger et al }(2008)}]{Schlamminger08}
\bibinfo{author}{S.~{Schlamminger et al }}, \bibinfo{journal}{Phys. Rev. Lett.}
  \bibinfo{volume}{100} (\bibinfo{year}{2008}) \bibinfo{pages}{041101}.
\bibitem[{{Sc{\'o}ccola et al}(2008)}]{Scoccola07}
\bibinfo{author}{C.G. {Sc{\'o}ccola et al}}, \bibinfo{journal}{Astrophys. J.}
  \bibinfo{volume}{681} (\bibinfo{year}{2008}) \bibinfo{pages}{737}.
\bibitem[{{Smith et al}(2000)}]{Smith00}
\bibinfo{author}{G.L. {Smith et al}}, \bibinfo{journal}{Phys. Rev. D}
  \bibinfo{volume}{61} (\bibinfo{year}{2000}) \bibinfo{pages}{022001}.
\bibitem[{{Srianand et al}(2004)}]{Srianand04}
\bibinfo{author}{R.~{Srianand et al}}, \bibinfo{journal}{Phys. Rev. Lett.}
  \bibinfo{volume}{92} (\bibinfo{year}{2004}) \bibinfo{pages}{121302}.
\bibitem[{{Su et al}(1994)}]{Su94}
\bibinfo{author}{Y.~{Su et al}}, \bibinfo{journal}{Phys. Rev. D}
  \bibinfo{volume}{50} (\bibinfo{year}{1994}) \bibinfo{pages}{3614}.
\bibitem[{{Sudarsky}(1992)}]{OscillatingBounds}
\bibinfo{author}{D.~{Sudarsky}}, \bibinfo{journal}{Phys. Lett. B}
  \bibinfo{volume}{281} (\bibinfo{year}{1992}) \bibinfo{pages}{98}.
\bibitem[{{Thompson et al}(2009)}]{Thompson09}
\bibinfo{author}{R.I. {Thompson et al}}, \bibinfo{journal}{Astrophys. J.}
  \bibinfo{volume}{703} (\bibinfo{year}{2009}) \bibinfo{pages}{1648}.
\bibitem[{{Webb et al }(2010)}]{Murphy10}
\bibinfo{author}{J.K. {Webb et al }}, \bibinfo{journal}{ArXiv e-prints
  1008.3907}  (\bibinfo{year}{2010}).
\bibitem[{{Webb et al}(1999)}]{Webb99}
\bibinfo{author}{J.K. {Webb et al}}, \bibinfo{journal}{Phys. Rev. Lett.}
  \bibinfo{volume}{82} (\bibinfo{year}{1999}) \bibinfo{pages}{884}.
\bibitem[{{Weinberg}(1983)}]{Weinberg83}
\bibinfo{author}{S.~{Weinberg}}, \bibinfo{journal}{Physics Letters B}
  \bibinfo{volume}{125} (\bibinfo{year}{1983}) \bibinfo{pages}{265}.
\bibitem[{{Wetterich}(2003)}]{Wetterich03}
\bibinfo{author}{C.~{Wetterich}}, \bibinfo{journal}{JCAP} \bibinfo{volume}{10}
  (\bibinfo{year}{2003}) \bibinfo{pages}{2}.
\bibitem[{{Wu} and {Wang}(1986)}]{Wu86}
\bibinfo{author}{Y.~{Wu}}, \bibinfo{author}{Z.~{Wang}}, \bibinfo{journal}{Phys.
  Rev. Lett.} \bibinfo{volume}{57} (\bibinfo{year}{1986})
  \bibinfo{pages}{1978}.
\bibitem[{{Youm}(2001{\natexlab{a}})}]{Youm2001a}
\bibinfo{author}{D.~{Youm}}, \bibinfo{journal}{Phys. Rev. D}
  \bibinfo{volume}{63} (\bibinfo{year}{2001}{\natexlab{a}})
  \bibinfo{pages}{125011}.
\bibitem[{{Youm}(2001{\natexlab{b}})}]{Youm2001b}
\bibinfo{author}{D.~{Youm}}, \bibinfo{journal}{Phys. Rev. D}
  \bibinfo{volume}{64} (\bibinfo{year}{2001}{\natexlab{b}})
  \bibinfo{pages}{085011}.

\end{thebibliography}
\bibliographystyle{model1b-num-names}

\end{document}